\documentclass[prd,twocolumn,showpacs,preprintnumbers,amsmath,amssymb]{revtex4}
\usepackage{amsmath}
\usepackage{amsfonts}
\usepackage{amssymb}
\usepackage{graphicx}
\usepackage{dcolumn}
\usepackage{bm}
\usepackage{color}
\newcommand{\im}{{\mbox{Im}\,}}

\newcommand{\ima}{\hbox{Im}\,}
\newcommand{\rea}{\hbox{Re}\,}

\newcommand{\sth}{s_{th}}
\newcommand{\dd}{\hbox{d}\,}
\newcommand{\imag}{{\mbox{Im}\,}}

\newcommand{\gammav}{ \Gamma }

\begin{document}

\title{Determination of $SU(2)$ Chiral Perturbation Theory low energy constants  
from a precise description of pion-pion scattering threshold parameters}

\author{J. Nebreda$^{a,b,c}$}
\author{J. R. Pel\'aez$^a$}
\author{G. R\'{\i}os$^{a,c,d}$}

\affiliation{$^a$Departamento de F{\'\i}sica Te{\'o}rica II,
  Universidad Complutense de Madrid, 28040   Madrid,\ \ Spain\\
$^b$ Yukawa Institute for Theoretical Physics, Kyoto University, 606-8502 Kyoto, Japan
\\ $^c$Helmholtz-Institut f\"ur Strahlen- und Kernphysik (Theorie) 
Universit\"at Bonn, D-53115 Bonn, Germany
\\
$^d$Departamento de F\'{\i}sica. Universidad de Murcia. E-30071, Murcia. Spain}
  
\begin{abstract}
We determine the values of the one- and two-loop low energy constants appearing
in the $SU(2)$ Chiral Perturbation Theory calculation of pion-pion scattering. For this
we use a recent and precise sum rule determination
 of some scattering lengths and 
slopes that appear in the effective range expansion. In addition we
provide sum rules for these coefficients
up to third order in the expansion.
Our results when using only the scattering lengths and slopes of the $S$, $P$, $D$ and $F$ waves
are consistent with previous determinations, but seem to require 
higher order contributions if they are to 
accommodate the third order coefficients of the 
effective range expansion.
\end{abstract}

\maketitle

\section{Introduction}

The smallness of the $u$ and $d$ quark masses together with
the spontaneous $SU(2)$ chiral symmetry breaking of QCD, which implies
 the existence of
the corresponding Goldstone bosons---the pions---allow us to write  
a low energy effective theory for QCD, organized as a systematic expansion
in pion masses and momenta.
This is known as Chiral Perturbation Theory (ChPT) \cite{chpt,Gasser:1983yg}, 
built as the most general low energy expansion of a Lagrangian containing just pions,
which is compatible with the 
symmetry constraints of QCD.
In particular, the first order is determined by
 the scale of the spontaneous symmetry breaking, 
identified with the pion decay constant $f_\pi$, and the pion mass $M_\pi$.
The expansion is then carried out in powers of $p^2/(4\pi f_\pi^2)$, where
$p$ denotes generically either the pion mass or momenta. 
The details of the QCD underlying dynamics are encoded in a set
of low energy constants (LECs), which
multiply the independent terms that appear in the Lagrangian
at higher orders. Note that all loop divergences appearing in a calculation up to 
a given order can be reabsorbed by renormalization of the
LECs up to that order.
In this process the LECs acquire a dependence on the renormalization scale $\mu$,
which is canceled with that present in the loops. In this way
calculations are rendered finite and scale independent to any given order of the expansion.

Only certain combinations of LECs appear in $\pi\pi$ scattering up to a given order.
As we commented above, to leading order $O(p^2)$ there are no LECs.
Within the $SU(2)$ formalism, to next-to-leading order (NLO), or $O(p^4)$, which corresponds to a one-loop calculation,
only four LECs, called $l_1, l_2,l_3$ and
$l_4$ appear in the amplitude, although two of them $l_3$ 
and $l_4$, do so through the quark mass dependence
of the pion mass $M_\pi$ and decay constant $f_\pi$. To next to next to leading order (NNLO), 
or $O(p^6)$,
six possible independent terms appear \cite{Knecht:1995tr}
multiplied by six constants, $\bar b_i$ with $i=1...6$,
which can be reexpanded in powers of the pion mass in terms of the four one-loop $l_k$
and six new NNLO LECs, denoted by $r_i$ \cite{Bijnens:1995yn}. After renormalization
these constants develop a scale dependence becoming $l_i^r(\mu)$ and $r^r_i(\mu)$, whereas
the $\bar b_i$ remain renormalization scale independent. 

Concerning the $O(p^4)$ LECs, we refer the reader to \cite{Colangelo:2010et}
for a recent compilation of lattice QCD and to \cite{Espriu:1989ff}
for some other estimates from quark-model-like calculations. It is also 
worth noticing that the bulk of their values can be explained by 
the effect of integrating out heavier resonances and, actually, seems to be saturated
by the vector multiplets \cite{Ecker:1988te}, plus a small contribution from scalars above 1 GeV, and an additional sizable contribution from kaons 
in the case of the SU(2) formalism
\cite{Gasser:1984gg}. Some estimates from resonance saturation have also been obtained 
for the $O(p^6)$ parameters \cite{Bijnens:1995yn}.
However, since perturbative QCD cannot be applied 
at very low energies, it is particularly difficult to obtain the values 
of these LECs from first principles and, with few exceptions, 
the LECs have been determined best from the comparison with
experiment \cite{Gasser:1983yg,Riggenbach:1990zp,Amoros:2000mc,Colangelo:2001df,Bijnens:2011tb}. 

Our aim in this work is to use a very recent dispersive analysis of data in 
\cite{GarciaMartin:2011cn},
which includes, among others, the latest very precise and reliable results on $K_{e4}$ decays from the
NA48/2 Collaboration~\cite{Batley:2010zza}, in order to determine the values of the $O(p^4)$ and $O(p^6)$ LECs that
appear in the $\pi\pi$ scattering amplitude. Since ChPT is a low energy amplitude,
obtained as a truncated series in powers of $(p/4\pi f_\pi)^{2n}$, we will compare
with data at threshold. In particular, we will obtain the LECs from fits 
to the coefficients of the momentum expansion of the amplitude around threshold, 
usually known as the effective range expansion. The coefficients of this expansion, even up to third order, are 
also becoming reachable for lattice calculations, although still only limited to the highest isospin channels
\cite{Beane:2011sc}.
 The most precise way to obtain these parameters
is by means of sum rules \cite{Wanders66,Olsson,BGN,ATW,AB1,AB2,GW1,GW2,ACGL,Colangelo:2001df,AB3} including 
those \cite{Palou:1974ma,Yndurain:2002ud,GarciaMartin:2011cn} 
obtained from the Froissart-Gribov representation \cite{FG,Yndurain:1972ix}, 
which we will use extensively here.
Actually, in several works the sum rules have already been used to determine 
the values of chiral parameters \cite{Gasser:1983yg,ATW,GW1,GW2,ACGL}. In addition, 
some bounds and constraints can be obtained from an axiomatic treatment 
\cite{axiomatic-oneloop,axiomatic-bs,axiomatic-ls-and-rs}.

Thus, after introducing the necessary notation,
 we present the experimental determination made in \cite{GarciaMartin:2011cn}, using sum rules, of the
threshold parameters up to second order of the effective range expansion.
In this work we consider one order more in that expansion, and we will determine, using the amplitudes in \cite{GarciaMartin:2011cn},
the values of the third order coefficients
up to the $F$ wave. To that end, we will use the existing Froissart-Gribov
sum rules \cite{Palou:1974ma}, but we will also derive here three sum rules
to calculate the third order threshold
parameters of the $S0$, $S2$ and $P$ waves with more precision. 
For convenience, we have grouped in an appendix all the sum rules used in the main text, explaining in detail how they are obtained and under what approximations they are related to other sum rules existing in the literature.
Some of the third order parameters are of relevance to obtain the values of the
LECs, since their leading-order contribution is directly proportional to 
combinations of LECs.  Actually, we carefully explain, for each threshold parameter, what 
is its leading order ChPT, and from what part of the calculation it stems from.
In Sec.~\ref{sec:op4fits} we first perform a fit of some of these parameters using just the $O(p^4)$ ChPT result, paying particular attention to an estimate of the systematic uncertainties in the parameters, which is of relevance for the role they will play in the determination of different LECs. Still, just by trying to describe a few threshold parameters, we are able to show
that the $O(p^4)$ approximation is not enough to describe the data at the present level of precision. In Sec.~\ref{sec:op6fits} we determine the best $\bar b_i$ constants and LECs that appear in the two-loop calculation. We will show that one can obtain 
 a relatively fair description in terms of $\bar b_i$ parameters, although the fact that the $\chi^2/{d.o.f.}$ 
of the fits is somewhat larger than one suggests that, at the present
level of precision, higher order contributions seem to be required.
In Sec.~\ref{sec:conclusions} we briefly discuss and summarize our results.

\section{Threshold parameters}

\subsection{Notation}
The amplitude for  $\pi\pi$ scattering is customarily decomposed 
in terms of partial waves $t^{I}_\ell$, 
of definite isospin $I$ and angular momentum $\ell$, as follows:
\begin{eqnarray}
  \label{eq:pwdecomp}
  F^{I}(s,t)&=&\frac{8}{\pi}\sum_{\ell}(2\ell+1)\, t^{I}_\ell(s)\, P_\ell(\cos\theta),\\
t^{I}_\ell(s)&=&\frac{1}{64\pi}\int_{-1}^1\, T^{I}(s,t,u) P_\ell(\cos\theta)\,d(\cos\theta),
\end{eqnarray}
$\theta$ 
being the scattering angle, $P_\ell$ the Legendre polynomials, $s,t,u$ 
the usual Mandelstam variables satisfying $s+t+u=4 M_\pi^2$,
and $T$ stands for the amplitude. 
In the elastic regime, the partial waves 
are uniquely determined by the 
phase shifts $\delta^{I}_\ell$ as follows:
\begin{equation}
  \label{eq:pwdef}
  t^{I}_\ell(s)=\frac{e^{i \delta^{I}_\ell(s)}\,\sin\delta^{I}_\ell(s)}{\sigma(s)},
\end{equation}
where $\sigma=2\,p/\sqrt{s}=\sqrt{1-4 M_\pi^2/s}$ and $p$ is the CM momentum. With this normalization, 
the effective range expansion of the real part of a partial wave can be written as
\begin{equation}
  \label{eq:effrange}
  \frac{1}{M_\pi} \rea t^{I}_\ell(s)= p^{2\ell}\Big( a_{\ell I}+b_{\ell I}\, p^2+\frac{1}{2}c_{\ell I} \,p^4+...\Big),
\end{equation}
where the $a_{\ell I}$ are usually called scattering lengths, the $b_{\ell I}$ slope parameters,
the $c_{\ell I}$ shape parameters,
and all of them, generically, threshold parameters.
Let us remark that it is usual to provide 
the values of these parameters {\it in units of $M_\pi$}, 
and we will do so in what follows. In addition, for odd waves we will drop
the isospin index, since it can only be $I=1$ due to Bose symmetry. Finally, we will also make use of the standard spectroscopic notation, where the $\ell=0,1,2,3...$ are called S, P, D, F... waves, followed
by the value of the isospin. 

\subsection{Structure of ChPT calculations}

\begin{table}
  \centering
  \begin{tabular}{c|c|c|c|c|c|c|}
 & $O(p^2)$ &\multicolumn{2}{c|}{ $O(p^4)$} & \multicolumn{3}{c|}{ $O(p^6)$}\\
 & pol. & $l_i$ pol. & $J$ & $r_i$ pol. & $l_i\,J$ & $ J^2, K$ \\
\hline
$a_S$&x&x& x & x & x&x \\ 
$b_S$&x&x& x & x & x&x \\ 
$c_S$& &x& x & x & x&x \\ 
$a_P$&x&x& x & x & x&x \\ 
$b_P$& &x& x & x & x&x \\ 
$c_P$& & & x & x & x&x \\ 
$a_D$& &x& x & x & x&x \\
$b_D$& & & x & x & x&x \\ 
$c_D$& & & x &   & x&x \\ 
$a_F$& & & x & x & x&x \\ 
$b_F$& & & x &   & x&x \\ 
$c_F$& & & x &   & x&x \\  
  \end{tabular}
  \caption{Contribution to threshold parameters from different 
orders and kinds of terms within ChPT, as explained in the text.
Recall that, due to Bose symmetry, those for $S$ and $D$ waves may have either isospin $0$ or $2$,
whereas those for $P$ and $F$ necessarily have isospin 1.}
\label{tab:abccontributions}
\end{table}

At this point it is relevant to 
discuss how the different orders of ChPT contribute to
each threshold parameter studied here, 
which we have gathered in the first column of
Table~\ref{tab:abccontributions}.
Let us start with the leading order, $O(p^2)$, of the ChPT amplitude,
which is a first order
polynomial in terms of Mandelstam variables $s,t,u$ and $M_\pi^2$. 
Since $s$ is independent
of $\theta$ whereas $t$ and $u$ are first order polynomials in
$\cos \theta$, the LO ChPT
can only contribute to the 
$a$ coefficients of the $S$ and $P$ waves and the
$b$ coefficients of the $S$ waves, but nothing to any other wave.
This corresponds to the second column in Table~\ref{tab:abccontributions}.
  
If we now consider the $O(p^4)$ amplitude, 
we find two kinds of terms. First, a polynomial, 
which includes
the $l_i(\mu)$ LECs and contains up to two powers of $\cos\theta$, 
so that it contributes
to the $a$ coefficients of $S$, $P$ and $D$ waves, the $b$ coefficients of 
the $S$ and $P$  waves as well as
the $c$ coefficients of the $S$ waves. 
This is the third column in Table~\ref{tab:abccontributions}.
However, there is another kind of 
$O(p^4)$ contributions, which come from the loop functions, called $J(q^2)$,
with two intermediate pions exchanged in any channel.
These loop functions carry a nonpolynomial 
dependence on $t$ and $u$ and 
therefore contribute to all waves, but note that they do not depend on $l_i(\mu)$. This is the fourth column, labeled ``$J$'', in Table~\ref{tab:abccontributions}.

Next, to two loops, $O(p^6)$, we find three kinds of terms:
First, pure polynomial terms containing the $r_i$ LECs,
which can contribute to
the $a$ coefficients up to the $F$ wave, $b$ coefficients up to $D$ waves, and $c$ coefficients
up to $P$ waves. These appear in the fifth column of Table~\ref{tab:abccontributions}.
In addition there are 
terms contributing to all waves, as shown in column six, containing
a single one-loop function and $l_i(\mu)$
LECs. Finally, there are also terms without LECs, which correspond to the last column,
containing either two one-loop functions
or one two-loop function, that we generically call $K(q^2)$.

Therefore, we see that only the $a_S, b_S$ and $a_P$ have a leading-order contribution
independent of the LECs. 
A priori, one could expect that the best observables to determine the $l_i$ are those 
whose leading term is $O(p^4)$, as it is the case of
the $a_{D0}$ and $a_{D2}$, which, actually, have been frequently used to determine the value of $l_1$ and $l_2$~\cite{Gasser:1983yg}.
Nevertheless, let us remark 
that there are still other threshold parameters whose leading contributions are
proportional to a combination of $l_i$, namely,  $b_P$, $c_{S0}$ and $c_{S2}$. However, 
these are much harder to determine reliably from experiment
and had not been used so far within the $S$, $P$ wave approximation for the absorptive part inside sum rules~\cite{AB2}. For this reason, in the next 
subsection we will explain how to obtain sum rules
for their determination.
To extract the $O(p^6)$ contributions and the $r_i$ LECs
is more difficult, not only because they are generically a smaller effect,
but also because they appear together with $O(p^4)$ terms or with $O(p^6)$ 
terms containing the $l_i$ and a one-loop function.
 
\subsection{Threshold parameters from sum rules}

The use of sum rules to obtain the values of threshold parameters is a well-established technique \cite{Wanders66,Olsson,Palou:1974ma,BGN,ATW,AB1,AB2,GW1,GW2,ACGL,Colangelo:2001df,AB3}. These can be obtained from different kinds of
dispersion relations with different numbers of subtractions. Two subtractions ensure the convergence of the dispersive integrals, but for certain channels fewer subtractions are also admissible. Actually, for fixed $t$ dispersion relations it is very convenient to work with symmetric or antisymmetric amplitudes under the $s\leftrightarrow u$ exchange.  Examples of these amplitudes are the symmetric $F^{00}$ and  $F^{0+}$ 
amplitudes corresponding to $\pi^0\pi^0\rightarrow\pi^0\pi^0$
and $\pi^0\pi^+\rightarrow\pi^0\pi^+$ amplitudes and the antisymmetric $F^{I_t=1}$ amplitude with isospin one in the $t$ channel.
Let us recall that in terms of definite isospin amplitudes in the $s$ channel $F^{0+}=F^{2}/2+F^{1}/2$, $F^{00}=2 F^{2}/3+F^{0}/3$ and $F^{I_t=1}=F^{0}/3+F^{1}/2-5F^{2}/6$.  Note also that the subtractions needed for the 
dispersion relation in the different scattering channels are not independent since, using crossing symmetry, it was shown by Roy \cite{Roy} that all them can be recast in terms of the $a_{S0}$ and $a_{S2}$ scattering lengths. Moreover, a forward $t=0$ dispersion relation for $F^{I_t=1}$ only needs one subtraction and, at threshold yields the well-known Olsson sum rule \cite{Olsson} that determines
the $2a_{S0}+5a_{S2}$ combination. However, a powerful set of sum rules for threshold parameters \cite{Palou:1974ma}, 
has also been obtained using the Froissart-Gribov representation \cite{FG} 
of the $t$ channel partial wave expansion of the antisymmetric $F^{I_t=1}$, 
and the symmetric $F^{0+}$ and $F^{00}$ waves. The resulting sum rules do not require subtractions for partial waves with $l\geq 1$ (for a pedagogical review see \cite{Yndurain:2002ud}). For completeness, we have collected all the sum rules used in this work in Appendix \ref{ap:sum rules}. Let us finally recall that in \cite{Mahoux:1974ej} it was shown that if one was to retain only the absorptive part of the $S$ and $P$ waves in twice-subtracted dispersion relations, the amplitude would be completely crossing symmetric and the dispersion relations of the Roy or Froissart-Gribov type, and their corresponding sum rules, would be identical. Actually, in Appendix C we will show how this is the case by recovering the
sum rules for $c$ parameters provided in \cite{AB2} starting from our sum rules under this approximation. Nevertheless, let us remark that in this work we will consider not only the $S$ and $P$ waves, but the $D$ and $F$ waves as well, and also that not all our sum rules are based on twice-subtracted dispersion relations.

One of the most important differences with previous determinations using sum rules is that we are going to use the recent, simple and very precise  data parametrizations obtained
in \cite{GarciaMartin:2011cn}. The relevance of those parametrizations is that they are obtained from data fits which have been highly constrained to satisfy three sets of dispersion relations within uncertainties: forward dispersion relations (FDRs) up to 1420 MeV, and Roy equations 
as well as once-subtracted Roy-like equations up to 1100 MeV. Above 1420 MeV, Regge expressions, assuming factorization, were fitted to $NN$, $N\pi$ and $\pi\pi$ total cross sections,
and used inside the integrals, allowing for the variation of the Regge parameters within the constrained fits to data. 
In that work, the values of the $a$ and $b$ threshold parameters were already provided
for the $S0$, $S2$, $P$, $D0$, $D2$ and $F$ waves, namely, all the combinations of $I=0,1,2$ and
$\ell=0,1,2,3$ allowed by Bose symmetry when considering pions as identical particles.
With the aim of minimizing the uncertainties, they were obtained from sum rules,
with the only exception of the $5 a_{S0}+2a_{S2}$ combination, which is orthogonal to the one appearing in the Olsson sum rule \cite{Olsson}. The results from sum rules were consistent with, but more accurate than, those directly obtained from the simple phenomenological parametrizations. Let us nevertheless remark that the sum rules used in \cite{GarciaMartin:2011cn} as well as those we will describe below have a very small dependence on the high energy region.  Actually, the high energy fits  used in \cite{GarciaMartin:2011cn} are just an updated version of the parametrizations of Regge behavior proposed in \cite{Pelaez:2003ky}, but other parametrizations exist \cite{Caprini:2011ky}. They have some differences, particularly for the uncertainties on the $t$ behavior, where no data exists, but the two of them overlap for forward scattering, which is the only one of relevance for the sum rules we will use here.

These results provided us with 12 observables determined from experiment,
which we list in Table~\ref{abcexp},
that we want to fit using four $l_i$ LECs in the 
one-loop case and six parameters $\bar b_i$ in the
two-loop case, which can be parametrized in terms of ten LECs.
Moreover, in  order to enlarge the set of observables that we include in our fit, we will also provide here
the calculation of the third order coefficient $c$ of the effective range expansion in 
Eq.\eqref{eq:effrange} above. These are five additional observables.

For this purpose, the Froissart-Gribov sum rules, used in \cite{GarciaMartin:2011cn} for scattering lengths and slopes,  
 allow us to write,  for $\ell>0$, the $c$ parameters as \cite{Palou:1974ma}
\begin{widetext}
\begin{eqnarray}
  \label{eq:FG}
  c_{\ell I}\!\!=\!\!\frac{\sqrt{\pi}\,\Gamma(\ell+1)}{M_\pi\,\Gamma(\ell+3/2)}
\!\int_{4M_\pi^2}^{\infty} \!\!\!\!ds
\left\{\frac{16 \,{\im F^{I}}''_{\!\!\!\!\!\cos\theta}(s,4M_\pi^2)}{(s-4M_\pi^2)^2 s^{\ell+1}} \right.
\!-\!8(\ell+1)\frac{\im {F^{I}}'_{\!\!\!\!\!\cos\theta}(s,4M_\pi^2)}{(s-4M_\pi^2)s^{\ell+2}}
\left.+\frac{\im F^{I}(s,4M_\pi^2)}{s^{\ell+3}} \frac{(\ell+2)^2(\ell+1)}{\ell+3/2}\right\},
\end{eqnarray}
\end{widetext}
where ${F^I}'_{\!\!\!\!\cos\theta}(s,4M_\pi^2)=(\partial/\partial\cos\theta) F^I(s,t)\vert_{t=4M_\pi^2}$ and $\theta$ is the angle between initial and final pions. 
These formulas allow us to calculate the $c$ parameters for the $P$, $D0$, $D2$ and $F$ wave,
that we list in Table~\ref{abcexp}. Note that the resulting values
 are all very  accurate with the exception of the  $c_P$ coefficient, and that the above sum rules
are not applicable to the scalar case.
These are the reasons why we provide here three new expressions of sum rules, one  for $c_P$ and two for $c_{S0}$ and $c_{S2}$,
\begin{widetext}
\begin{equation}
c_P=-\frac{14 \,a_F}{3}+\frac{16}{3 M_\pi}
\int_{4M_\pi^2}^{\infty} ds
\left\{ \frac{\ima F^{0}(s,0)}{3s^4}-
\frac{\ima F^{1}(s,0)}{2s^4}-\frac{5\ima F^{2}(s,0)}{6s^4} 
+\left[\frac{\ima F^{1}(s,0)}{(s-4M_\pi^2)^4}-\frac{3 a_P^2 M_\pi }{4\pi(s-4M_\pi^2)^{3/2}}\right]\right\},
\label{eq:newSRcp} 
\end{equation}
\begin{eqnarray}
&&c_{S2}=-6b_P-10a_{D2}+\frac{8}{M_\pi}\int_{4M_\pi^2}^\infty ds \left\{\frac{\ima F^{0+}(s,0)}{s^3}+\frac{1}{(s-4M_\pi^2)^{5/2}}\right.\\
&&\hspace{6.cm}\times\left.\left[\frac{\ima F^{0+}(s,0)}{\sqrt{s-4M_\pi^2}}
-\frac{2M_\pi a_{S2}^2}{\pi}-
\frac{s-4M_\pi^2}{\pi}
\left(\frac{M_\pi}{2}(2 a_{S2}b_{S2}+a_{S2}^4)-\frac{a_{S2}^2}{4M_\pi}\right)
\right]
\right\},\nonumber
\end{eqnarray}
\begin{eqnarray} 
&&c_{S0}=-2c_{S2}-20a_{D2}-10a_{D0}+\frac{12}{M_\pi}\int_{4M_\pi^2}^\infty ds \left\{\frac{\ima F^{00}(s,0)}{s^3}+\frac{1}{(s-4M_\pi^2)^{5/2}}\right.\\
&&\label{eq:newSRcS0}\hspace{.5cm}\times\left.\left[\frac{\ima F^{00}(s,0)}{\sqrt{s-4M_\pi^2}}
-\frac{4M_\pi (2a_{S2}^2+a_{S0}^2)}{3\pi}-
\frac{s-4M_\pi^2}{3\pi}
\left(M_\pi[2 (2a_{S2}b_{S2}+a_{S2}^4)+2a_{S0}b_{S0}+a_{S0}^4]-\frac{2a_{S2}^2+a_{S0}^2}{2M_\pi}\right)
\right]
\right\}.\nonumber
\end{eqnarray}
\end{widetext}

Let us note that, in all these sum rules, we have written several terms together 
inside square brackets to emphasize that they do not converge separately.
 The derivation is similar to 
the sum rules obtained for $b_P$, $b_{S0}$ and $b_{S2}$ in \cite{PY05,KPY08}. They correspond to the 
threshold limit, taken from above,
 of the second derivative of a forward dispersion relation for the $F^{I_t=1}$, $F^{0+}$ and $F^{00}$ amplitudes, 
 respectively. In Appendix C we list them again together with those used
for scattering lengths and slopes, but this time, for completeness, in terms of $F^I$ amplitudes, more convenient for calculations,
instead of the $F^{00}$, $F^{0+}$ and $F^{I_t=1}$ used for the original derivation.

In the above sum rules we have explicitly got rid of the principal part (P.P.) that appears in the 
dispersion relation by using:
\begin{equation}
{\rm P.P.}\int_{0}^{\infty}\frac{dx}{(x-y)\sqrt{x}}=0,  \,{\rm for}\,  y >0.
\end{equation}

As we will explain in detail in the Appendix, {\it before the principal part has been removed from the sum rules with this trick}, if the threshold parameters outside the 
integrals are replaced by their Froissart-Gribov sum rules (given in Appendix \ref{appendix:FG}) and at the same time the
 absorptive parts inside the integrals are approximated by just the $S$ and $P$ partial wave contributions, one recovers the sum rules for the $c$ parameters obtained in \cite{AB2}, 
also {\it before the principal part is removed
from their integrals}. The trick used in \cite{AB2} to remove the principal part is similar but not the same as ours. Let us nevertheless remark once more that our sum rules above contain in principle all contributions from all partial waves, not just $S$ and $P$.

\begin{table}
  \centering
  \renewcommand{\arraystretch}{1.25}
  \begin{tabular}{l|c|c|c}
& CFD & Sum rules & Best value \\
\hline
$a_{S0}$               &  $0.221\pm0.009$  &                    & {\bf$0.220\pm0.008$}\\
$a_{S2}(\times 10^2)$  &  $-4.3\pm0.8$ &                        & {\bf $-4.2\pm0.4$}\\
$2a_{S0}-5 a_{S2}$     &  $0.657\pm0.043$  & $0.648\pm0.016$    & {\bf $0.650\pm0.015$}\\
$a_{P}$($\times 10^3$) &  $38.5\pm1.2$     & $37.7\pm1.3$       & {\bf $38.1\pm0.9$}\\
$a_{D0}$($\times 10^4$)&  $18.8\pm0.4$     & $17.8\pm0.3$       & {\bf $17.8\pm0.3$}\\
$a_{D2}$($\times 10^4$)&  $2.8\pm1.0$      & $1.85\pm0.18$      & {\bf $1.85\pm0.18$}\\
$a_{F}$($\times 10^5$) &  $5.1\pm1.3$      & $5.65\pm0.23$      & {\bf $5.65\pm0.23$}\\ \hline
$b_{S0}$               &  $0.278\pm0.007$  & $0.278\pm0.008$    & {\bf $0.278\pm0.005$}\\
$b_{S2}(\times 10^2)$  &  $-8.0\pm0.9$ & $-8.2\pm0.4$   & {\bf $-8.2\pm0.4$}\\
$b_{P}$($\times 10^3$) &  $5.07\pm0.26$    & {\small $6.0\pm0.9$,$5.48\pm0.17$}& {\bf $5.37\pm0.14$}\\
$b_{D0}$($\times 10^4$)&  $-4.2\pm0.3$     & $-3.5\pm0.2$       & {\bf $-3.5\pm0.2$}\\
$b_{D2}$($\times 10^4$)&  $-2.8\pm0.8$     & $-3.3\pm0.1$       & {\bf $-3.3\pm0.1$} \\
$b_{F}$($\times 10^5$) &  $-4.6\pm2.5$     & $-4.06\pm0.27$     & {\bf $-4.06\pm0.27$}\\\hline
$c_{S0}(\times 10^2)$  &  $-0.12\pm1.22$   & $0.7\pm0.8$        & {\bf $0.45\pm0.67$    }\\
$c_{S2}(\times 10^2)$  &  $3.6\pm1.8$      & $2.79\pm0.24$      & {\bf $2.80\pm0.24$} \\
$c_{P}$($\times 10^3$) &  $1.41\pm0.19$    & $2.3\pm0.8$,$1.35\pm0.15$         & {\bf $1.39\pm0.12$}\\
$c_{D0}$($\times 10^4$) &  $5.6\pm0.4$      & $4.4\pm0.3$        & {\bf  $4.4\pm0.3$}\\
$c_{D2}$($\times 10^4$) &  $5.5\pm1.6$      & $3.6\pm0.2$        & {\bf $3.6\pm0.2$}\\
$c_{F}$($\times 10^5$) &  $11\pm9$         & $6.9\pm0.4$        & {\bf $6.9\pm0.4$}\\
  \end{tabular}
  \caption{Values of threshold parameters obtained in Ref.~\cite{GarciaMartin:2011cn}
together with those obtained here for the $c$ parameters. The ``CFD'' column lists the values as obtained directly from the "Constrained Fits to Data" provided in 
\cite{GarciaMartin:2011cn}. 
We also provide the values obtained from sum rules. We typically consider these our best results,
except in cases when the CFD  are competitive and not very correlated with the sum rule. Note that for $b_P$ and $c_P$ waves we provide two values. For $c_P$, the first one, less accurate, corresponds to the Froissart-Gribov sum rule 
in Eq.\eqref{eq:FG} and the second one to the sum rule in Eq.\eqref{eq:newSRcp}.
Similarly, for $b_P$, the first, less precise result, is from the Froissart-Gribov sum rule, and the second from a fast convergent sum rule, as explained in \cite{PY05,KPY08}. 
\label{abcexp}}
\end{table}

At this point, a comment about isospin breaking is in order. The whole formalism we have described so far, either for sum rules or ChPT is isospin symmetric; namely, we have set $m_u=m_d$ and neglected electromagnetic effects. Thus, customarily all pion masses are set to the charged one, 
and all of them have a single decay constant $f_\pi=92.4$ MeV. In the literature this scenario has been sometimes referred to as a ``paradise world'' \cite{Gasser:2007de}, and is also the standard one in previous sum rule determinations \cite{Wanders66,Olsson,BGN,ATW,AB1,AB2,GW1,GW2,ACGL,Colangelo:2001df,AB3,Palou:1974ma,Yndurain:2002ud,GarciaMartin:2011cn,FG,Yndurain:1972ix,Gasser:1983yg,axiomatic-oneloop,axiomatic-bs,axiomatic-ls-and-rs}. Of course, experimental data are obtained in the real world and the isospin-breaking effects have to be subtracted from the data or considered as an additional source of uncertainty. In particular, there is an spectacular isospin-breaking effect appearing in the scalar-isoscalar phase shifts close to threshold, which are obtained from $K_{\ell4}$ decays, enhanced over the typical expectations due to the proximity of the different $\pi^0\pi^0$ and $\pi^+\pi^-$ thresholds. This effect is not present in other channels, like the $P$ wave, or isospin 2, etc. 
In this work this enhanced effect has been properly subtracted, since
 the dispersive parametrizations that we have taken from \cite{GarciaMartin:2011cn} use the phase shifts from NA48/2 and other $K_{\ell4}$ experiments, but corrected
from this isospin-breaking effect following the formalism obtained in \cite{Gasser:2007de} (these corrections were also obtained in \cite{Cuplov:2003bj}). 

Once this near threshold isospin-breaking enhancement has been accounted for, one might wonder about the typical size of isospin-breaking corrections on the phase shifts at any energy for all waves, that one would expect to lie below 3\% due to the difference between the charged and neutral pion masses. There is no calculation available to subtract this effect and obtain the isospin symmetric phases from experiments. Thus, it is customary to consider this as part of the uncertainty in the experimental input. In the case of the data fits that we use from \cite{GarciaMartin:2011cn}, the uncertainties used
as {\it input} experimental data are either much larger (by factors of 3 to 4) than  3\%  or, as in the vector channel, include a systematic uncertainty obtained as the difference between data parametrizations with different pion masses \cite{PY05}. Thus this effect is part of the input uncertainties and propagates to the uncertainties of the isospin symmetric final results. Other dispersive approaches as, for instance, that in \cite{Colangelo:2001df} also use as input (for their matching point, for other partial waves or for the high energy amplitudes) experimental values whose large uncertainties cover well the expected contribution from isospin-breaking effects.

One could also be interested in obtaining the scattering lengths in the presence of isospin breaking. The corresponding ChPT expressions needed to include these effects and obtain the scattering lengths for the different mass channels have been worked out in \cite{isospinbreaking}. These of course, need the introduction of some other low energy constants. This calculation lies beyond the scope of this work, where we only focus in the traditional ``paradise world'' isospin symmetric formalism.

\section{$O(p^4)$ fits}
\label{sec:op4fits}

Before presenting the fits to the full two-loop ChPT results, it is instructive to
try to fit the threshold parameters by means of the one-loop ChPT amplitudes.
This will help us check the stability of the LECs values
and the need for higher order counterterms, but it will also help us illustrate
our different fitting strategies in order to deal with systematic uncertainties.

Let us recall once more that to $O(p^4)$ only four LECs, appear in $\pi\pi$ scattering,
customarily denoted by $\bar l_1,...,\bar l_4$, which are basically the $l^r_i(\mu)$
at the $\mu=M_\pi$ scale and normalized so that they have values of order 
one~\cite{Gasser:1983yg}.
Note, however, that $\bar l_3$ and $\bar l_4$ only appear through the 
quark mass dependence of $M_\pi$ and $f_\pi$, respectively, and therefore we 
cannot expect much sensitivity to these two parameters from fits to the coefficients
of the momentum expansion of amplitudes.

In addition, from Table~\ref{tab:abccontributions}, we see that, up to $O(p^4)$,
 only ten observables carry any dependence on the LECs: for
half of them, $a_{S0}$, $a_{S2}$, $a_P$, $b_{S0}$ and $b_{S2}$,  the leading contribution
is $O(p^2)$, whereas for the other five the leading contribution is directly of $O(p^4)$.
Therefore, we expect the latter to be more sensitive to the LECs, 
but also to the higher order corrections that we are neglecting.

Thus, in Table~\ref{fits-op4} we show the results of  our fits.
 First, we have fitted only  the observables
which have a leading $O(p^2)$ contribution,  since, in principle these might be
more stable under the higher order corrections. The fit comes out with relatively low 
$\chi^2/d.o.f.$ Next we have presented a determination of $\bar l_1$ and $\bar l_2$,
which are, in principle, fixed from $a_{D0}$ 
and $a_{D2}$ alone, which are not included in the previous fit.
It is evident that the resulting values from those two fits are incompatible, 
particularly $\bar l_1$.  The same happens when we determine their values from $c_{S0}$ and $c_{S2}$ alone.  The incompatibility is even worse when fitting simultaneously 
the ten observables that depend on $\bar l_i$ to $O(p^4)$. 
These results imply that, as is well known, to the present level of precision 
the one-loop ChPT formalism is not enough and calls for higher order corrections.

\begin{table*}[t]
 \centering
  \renewcommand{\arraystretch}{1.25}
  \begin{center}
    \begin{tabular*}{0,8\textwidth}{@{\extracolsep{\fill} }lccccc}
      \hline \hline
      Fit to& $\bar l_1$ &	 $\bar l_2$ & $\bar l_3$ & $\bar l_4$ & $\chi^2/d.o.f.$ \\
      \hline
      $a_S, b_S, a_P$ & $1.1\pm1.0$ &  $ 5.1\pm0.7$ &  $-1\pm8$ &   $7.1\pm0.7$ &
      0.23 \\
      $a_D$ & $-1.75\pm0.22$ & $5.91\pm0.10$ & --- & --- & 0  \\
      $c_S$   & $-2.4\pm0.9$ &   $4.8\pm0.4$ & --- &  --- & 
      0  \\ 
      $a_S, b_S, a_P,a_D,c_S,b_P$ &  $-2.06\pm0.14$ &  $ 5.97\pm0.07 $&  $-5\pm 8$ &   $7.1\pm0.6$ &
      7.9  \\ 
      $a_S, b_S, a_P,a_D,c_S,b_P$,  using $f_0$   &  $-1.06\pm0.11$ &  $ 4.6\pm0.9$ &  $0\pm6$ &  $ 5.0\pm0.3$ &
    $7.06$  \\ 
      \hline\hline
   {\bf Our  Estimate $\mathbf{O(p^4)}$  }&  $\mathbf{ -1.5\pm0.5}$  &  {$\mathbf {5.3\pm0.7}$  }&  {$\mathbf {-3\pm7}$  }&  {$\mathbf {6.0\pm1.2} $ }& ---  \\ \hline\hline
      \hline
    \end{tabular*}
  \end{center}
  \caption{$O(p^4)$ fits to different sets of threshold parameters containing polynomial $O(p^4)$ contributions. Note that the results of the three first lines are rather incompatible with each other. This is also illustrated by the large $\chi^2/d.o.f.$ 
  when fitting all the observables simultaneously. We also show two versions of such a fit, either using $f_\pi$ or $f_0$ in the last order of the expansion. Finally, we provide an estimate of how much one should enlarge the uncertainties of the LECs if, for simplicity, one still insists in using the one-loop formalism. Beyond that accuracy a two-loop formalism is called for.}
  \label{fits-op4}
\end{table*}

\begin{table*}[t]
 \centering
  \renewcommand{\arraystretch}{1.25}
  \begin{center}
    \begin{tabular*}{0,8\textwidth}{@{\extracolsep{\fill} }lcccc}
      \hline \hline
        & $\bar l_1$ & $\bar l_2$ & $\bar l_3$ & $\bar l_4$  \\
      \hline
      Numerical analysis \\
      \hline
	Ref.~\cite{Colangelo:2001df} ``matching at one-loop"& $-1.8$ & 5.4 & --- & --- \\
	Ref.~\cite{Colangelo:2001df}& $-0.4\pm0.6$ & $4.3\pm0.1$ & --- & $4.4\pm0.2$ \\
	Ref.~\cite{Bijnens:2011tb} ``All NLO''& 1.1 & 4.6 & 4.9 & 4.8\\
	Ref.~\cite{Bijnens:2011tb} ``All NNLO''& -0.1 & 5.3 & 4.2 & 4.8 \\
      \hline
      Lattice analysis\\
      \hline
      Ref.~\cite{milc09a} SU(2) fit & --- & --- & $3.0\pm0.6^{+0.9}_{-0.6}$ & $3.9\pm0.2\pm0.3$\\      
      Ref.~\cite{milc10a} & --- & --- & $2.85\pm0.81^{+0.37}_{-0.92}$ & $3.98\pm0.32^{+0.51}_{-0.28}$\\
      Ref.~\cite{RBC-UKCD}& --- & --- & $2.57\pm0.18$ & $3.83\pm0.09$\\      
      Ref.~\cite{JLQCD}& \multicolumn{2}{c}{$\bar l_1 - \bar l_2=-2.9\pm0.9\pm1.3$} & --- & $4.09\pm0.50\pm0.52$   \\      
      Ref.~\cite{Colangelo:2010et} & --- & --- & $3.2\pm0.8$ & --- \\
      \hline
    \end{tabular*}
  \end{center}
  \caption{Different determinations of the $O(p^4)$ shown for comparison with our results. The upper section of the table shows some phenomenological determinations~\cite{Colangelo:2001df,Bijnens:2011tb} and the lower section shows several determinations by different lattice groups~\cite{milc09a,milc10a,RBC-UKCD,JLQCD} and by the Flavianet Lattice Averaging Group FLAG~\cite{Colangelo:2010et}. }
  \label{comparison-lecs-Op4}
\end{table*}

For instance, the effect of these higher order corrections can be seen by fitting
to the one-loop amplitude but replacing $f_\pi$ by $f_0$ in the $O(p^4)$ terms,
since the resulting expression is also correct up to $O(p^4)$, 
only differing in higher order contributions. This we show in row 5 of Table~\ref{fits-op4}.
Surprisingly, the $\chi^2/d.o.f.$ comes somewhat lower, but the values of the LECs
come out rather different from the previous calculation. 

Of course, if one still wants to use the 
relatively simple $O(p^4)$  approximation instead of the full two-loop amplitude, one could always try to include the effect of higher orders 
into a systematic uncertainty of the LECs, 
at the expense of accuracy. In such case we propose to take the 
weighted average of the two previous fits, including a systematic 
uncertainty to cover the LECs
values of both fits \footnote{We have weighted each fit by the square of its $\chi^2/d.o.f.$, whereas the uncertainty is obtained by adding in quadrature the averaged statistical uncertainty to the maximum difference between the resulting central value and the central value of each fit.}. 
This corresponds to the values in the ``Our Estimate $O(p^4)$'' row in Table~\ref{fits-op4}.

For comparison, in Table~\ref{comparison-lecs-Op4}, we have included other sets available in the literature together with several recent lattice estimates.  Actually,
our results are pretty close to those obtained in \cite{Colangelo:2001df}, particularly
to the ``matching at one-loop" set. Moreover, we can compare with the SU(2) parameters translated from the $SU(3)$ LECs which were obtained in \cite{Bijnens:2011tb} by using NLO and NNLO SU(3) ChPT to fit many observables like scattering lengths and slope parameters for $\pi K$, $\pi\pi$ scattering (including the latest NA48/2 data), form factors, the $m_s/\hat m$ quark mass ratio, etc. Their NLO result is very close to our fit to $a_S,b_S,a_P$, particularly for $\bar l_1$, but their NNLO result gets closer to our final estimate here.

In Table~\ref{tab:our-estimate} we compare 
the resulting threshold parameters obtained
using this averaged set with the ``Best value" of Table~\ref{abcexp}, which we repeat under the
``Data analysis" column.
\begin{table}[b]
  \centering
    \renewcommand{\arraystretch}{1.25}
  \begin{tabular}{c|cc|c}
    & Estimate & Estimate  & Data  \\
    & $O(p^4)$ & $O(p^6)$  & Analysis \\\hline
 $a_{S0}$               &     0.214$\pm$0.009 & 0.230 $\pm$0.014 & {\bf$0.220\pm0.008$}\\
 $a_{S2}(\times 10^2)$  &    -4.4$\pm$0.3  &-4.3$\pm$0.4   & {\bf $-4.2\pm0.4$}\\ 
 $a_{P}$($\times 10^3$) &     38.7$\pm$1.2 & 39.0$\pm$0.8      & {\bf $38.1\pm0.9$}\\
 $a_{D0}$($\times 10^4$)&     15$\pm$3      & 16.9$\pm$0.9    & {\bf $17.8\pm0.3$}\\
 $a_{D2}$($\times 10^4$)&     1.3$\pm$1.0   & 1.7$\pm$0.3    & {\bf $1.85\pm0.18$}\\
 $a_{F}$($\times 10^5$) & ---               & 4.6$\pm$0.5    & {\bf $5.65\pm0.23$}\\ 
 \hline
 $b_{S0}$               &   0.255$\pm$0.011 & 0.271$\pm$0.007 & {\bf $0.278\pm0.005$}\\
 $b_{S2}(\times 10^2)$  &   -8.2$\pm$0.5    &-8.4$\pm$0.2    & {\bf $-8.2\pm0.4$}\\
 $b_{P}$($\times 10^3$) &      4.4$\pm$0.5  &  5.2$\pm$0.2    & {\bf $5.37\pm0.14$}\\
 $b_{D0}$($\times 10^4$) & ---              & -3.6$\pm$0.8     & {\bf $-3.5\pm0.2$}\\
 $b_{D2}$($\times 10^4$) & ---              & -3.1$\pm$0.4     & {\bf $-3.3\pm0.1$} \\
 $b_{F}$($\times 10^5$) & ---               & -3.4$\pm$0.3& {\bf $-4.06\pm0.27$}\\
 \hline
 $c_{S0}(\times 10^2)$   &  2.3$\pm$1.4     &  1.3$\pm$0.6     & {\bf $0.45\pm0.67$}\\
 $c_{S2}(\times 10^2)$   &  3.4$\pm$0.7     &  2.78$\pm$0.16   & {\bf $2.80\pm0.24$} \\
 $c_{P}$($\times 10^3$)  & ---               &  0.3$\pm$0.2    & {\bf $1.39\pm0.12$}\\
 $c_{D0}$($\times 10^4$) & ---               &  3.6$\pm$0.2    & {\bf  $4.4\pm0.3$}\\
 $c_{D2}$($\times 10^4$) & ---               &  3.2$\pm$0.2    & {\bf $3.6\pm0.2$}\\
 $c_{F}$($\times 10^5$) & ---               &  5.4$\pm$0.4    & {\bf $6.9\pm0.4$}\\
  \end{tabular}
  \caption{Values of the threshold parameters obtained from $O(p^4)$ ChPT
  and $O(p^6)$ ChPT using the averaged sets of LECs from the sixth row of
  Table~\ref{fits-op4} and the fifth row of Table~\ref{tab:fullbs}. We
  also show in the third column the best values obtained from the data analysis
  given in Table~\ref{abcexp}.}
  \label{tab:our-estimate}
\end{table}
We can see there that, thanks to the larger uncertainty, the threshold parameters obtained are compatible within errors
with the experimental values, except for $b_{S0}$ and $b_P$, which differ by
more than 3 and 2 standard deviations respectively. Furthermore, we have explicitly checked that the LECs values in the ``Our Estimate $O(p^4)$'' set satisfy very comfortably the axiomatic constraints derived in \cite{axiomatic-oneloop}\footnote{For the constraints obtained within $SU(3)$ ChPT we have used the conversion between $SU(3)$ and $SU(2)$ LECs given in \cite{Gasser:1984gg}.}.

\section{$O(p^6)$ fits}
\label{sec:op6fits}
As we already commented in the Introduction, 
the threshold parameters
can be described in ChPT at $O(p^6)$ in terms of six low energy constants, usually denoted 
$\bar b_1,\cdots, \bar b_6$.  Let us remark, however, that the first four can be separated in two parts with different chiral order, namely,
$\bar b_i=\bar b_i^{(0)}+\Delta \bar b_i$, $i=1,2,3,4$, where
$\bar b_i^{(0)}=O(m_\pi^0)$ and $\Delta\bar b_i=O(m_\pi^2)$. The $\bar b^{(0)}_i$
parameters contain combinations of the four $O(p^4)$ LECs $\bar l_i$, 
but not of the $O(p^6)$ LECs. In contrast, six linear combinations of the latter appear inside the
$\Delta\bar b_i$ for $i=1...4$ as well as in $\bar b_5$ and $\bar b_6$, and are 
accordingly 
denoted by $r_i$, with $i=1...6$.
Due to this $O(m_\pi^2)$ part in the parameters, 
the calculations using the $\bar b_i$  have an extra $O(p^8)$
piece which is not present when using $\bar l_i$ and $r_i$
(or making the separation $\bar b_i=\bar b^{(0)}_i+\Delta\bar b_i$ explicit).
Of course, since this is a higher order
contribution, both descriptions are formally equivalent up to $O(p^6)$.  Nevertheless, there could be relevant numerical differences
and, what is more important to us, in one case one should determine
only 6 parameters, whereas in the other case there are 10 parameters.

Thus, when using $\bar l_i$ and $r_i$, we may obtain spurious solutions or, 
in general, less stable values than when using just the six $\bar b_i$. 
That is why in this section we have decided to use the $\bar b_i$ set in our fits. For completeness we provide in Appendix~\ref{subsec:ls&rs} a detailed account of our results when 
parametrizing the ChPT series in terms of
$\bar l_i$ and $r_i$.

\begin{table*}[t]
  \centering
    \renewcommand{\arraystretch}{1.25}
  \begin{tabular}{lccccccc}
    \hline \hline
    Fit to& $\bar b_1$ & $\bar b_2$& $\bar b_3$& $\bar b_4$& $\bar b_5$& $\bar b_6$&
    $\chi^2/d.o.f.$ \\
    \hline
    $a_S, b_S, a_P,a_D,c_S,b_P$ & -14$\pm$4 &   14.6$\pm$1.2 &  -0.29$\pm$0.05 & 
    0.76$\pm$0.02 &   0.1$\pm$1.1 &   2.2$\pm$0.2 & 6.0/(10-6+1)=1.2 \\ 
    All & -2$\pm$3 &    14.2$\pm$1.0 &  -0.39$\pm$0.04 & 0.746$\pm$0.013 &
    3.1$\pm$0.3 &    2.58$\pm$0.12 & 67/(18-6+1)=5.2   \\ 
    All but $c_P$  & -6$\pm$3 & 15.9$\pm$1.0 & -0.36$\pm$0.04 & 0.753$\pm$0.013 &
    2.2$\pm$0.4 &  2.44$\pm$0.12 &  34.9/(17-6+1)=2.9  \\
    All but $c_P$, using $f_0$ & -12$\pm$3 & 13.9$\pm$0.9 & -0.30$\pm$0.04 &  
    0.726$\pm$0.013 &  1.0$\pm$0.3 &  1.93$\pm$0.08 &  12.5/(17-6+1)=1.04  \\
    \hline\hline
     {\bf Our Estimate $O(p^6)$ } &  {\bf -10.5$\pm$5.1 }  & {\bf  14.5$\pm$1.8  } & {\bf  -0.31$\pm$0.06  } & {\bf  0.73$\pm$0.02  } &  {\bf 1.3$\pm$1.0  } &
    {\bf  2.1$\pm$0.4  } & --- \\  
    \hline\hline
    Ref.~\cite{Colangelo:2001df} & -12.4$\pm$1.6  & 11.8$\pm$0.6 & -0.33$\pm$0.07 & 0.74$\pm$0.01 & 3.6$\pm$0.4 & 2.35$\pm$0.02 \\
  \end{tabular}
    \caption{$O(p^6)$ fits to different sets of threshold parameters. In the first row we only fit to observables containing polynomial $O(p^4)$ contributions. Note the improvement of the
    $O(p^6)$ description versus the $O(p^4)$ one by comparing the $\chi^2/d.o.f.$ here with
    the corresponding one in Table~\ref{fits-op4}. Next we show the results of the fit to all the
    threshold parameters obtained in this work. Note that the fit quality is rather poor. However, most of
    the disagreement is caused by a single observable $c_P$. When this observable is omitted, the resulting fits are of much better quality, particularly good when using $f_0$ instead of $f_\pi$ in the last term of the ChPT expansion. We also provide an estimate of the LECs uncertainties from the fits to all observables except $c_P$, as a weighted average of the fits using $f_0$ or $f_\pi$. Within our uncertainties, the resulting values of the $\bar b_i$ parameters
    are very consistent with previous determinations, listed in the last row. Let us remark that, as emphasized in \cite{Colangelo:2001df}, ``the error bars only indicate the noise'' seen in their evaluation. This would correspond to our uncertainties in the ``All but $c_P$, using $f_0$'' row, whereas the error bars we provide 
for our final result also contain some crude estimate of higher order uncertainties. 
  }
    \label{tab:fullbs}
  \end{table*}

Therefore, and similarly to the $O(p^4)$ case, we first fit the ten threshold parameters $a_S$,  $a_P$, $a_D$, $b_{S}$,
$b_P$, and $c_{S}$, which have
a nonzero $O(p^4)$ polynomial contribution, since we expect
these to be more stable under higher order ChPT corrections.
In the first row of 
Table~\ref{tab:fullbs} we show the resulting $\bar b_i$
for this fit, which describes fairly well the  fitted observables
with a $\chi^2/d.o.f.= 1.2$. However, the threshold parameters that are  not fitted,
are not so well described with this set of LECs. 

Actually, when fitting all 18 threshold parameters, we obtain somewhat different 
LECs, which are shown in the second row of Table~\ref{tab:fullbs}. 
Although not dramatically incompatible with those of the first row, we see
differences around two standard deviations for $\bar b_3$ and $\bar b_4$
and around 3 standard deviations for $\bar b_1$ and $\bar b_5$. 
Unfortunately, this second fit comes out
with a rather poor $\chi^2/d.o.f.=5.2$.
Therefore, it seems that we cannot describe all observables simultaneously 
with two-loop ChPT within the present level of precision. Higher order contributions
seem to be required.

Nevertheless, we have noticed that  $c_P$  alone contributes almost to one-third of the total $\chi^2$. This might indicate that $c_P$
receives 
important higher order contributions that are not being taken
into account in the $O(p^6)$ calculation.
Once again we can obtain a crude
estimate of the size of higher order ChPT corrections, by changing $f_\pi$ by $f_0$ in
the last term of the ChPT expansion. By far it is $c_P$ the one that suffers the
largest change, by almost $80\%$. Certainly it looks like a good candidate to receive
very large higher order ChPT corrections. 

Thus, we proceed to fit again all threshold parameters except
$c_P$, and the result is shown in the third row of Table~\ref{tab:fullbs}.
The fit quality improves considerably, but we still get a high $\chi^2/d.o.f.=2.9$, which
indicates that the two-loop calculation may not be enough to describe
even the remaining threshold parameters with their current
level of precision.

However, as a final attempt,  we can see the effect of higher order corrections by
making a fit replacing $f_\pi$ by $f_0$ in the $O(p^6)$ terms, 
since the resulting expression is also correct up to $O(p^6)$. 
We show the results (without including $c_P$) in the fourth row of Table~\ref{tab:fullbs}. 
Surprisingly, we now obtain a good $\chi^2/d.o.f.=1.0$, and all LECs are
less than 2 standard deviations from those obtained by  fitting only the
threshold parameters which have an $O(p^4)$ polynomial part. 
We therefore conclude that, by excluding $c_P$, the two-loop fit
 still shows some tension but by conveniently
using $f_0$ the last term of the ChPT expansion, it can give an acceptable description
of the rest of the threshold parameters.

For this reason, we have once more
made a weighted average of the two fits 
(the one using $f_\pi$ and the one using $f_0$) adding systematic uncertainties to 
cover both sets. This we show in the fifth row of Table~\ref{tab:fullbs}, called ``Estimate $O(p^6)$''.

In addition, it can be noticed that ``Our Estimate $O(p^6)$'' set is very compatible with the results in \cite{Colangelo:2001df,Colangelo:2010et}. Note that four out of the six $b_i$ lie within the uncertainties, whereas only one of them, $b_5$ lies 2 deviations apart.  Actually, the agreement is even better than it may look at first sight, because, as emphasized in \cite{Colangelo:2001df}, ``the error bars only indicate the noise'' seen in their evaluation, and do not include effects from other sources of uncertainty. This would correspond to our uncertainties in the ``All but $c_P$, using $f_0$'' row, whereas the error bars we provide 
for our final result also contain some crude estimate of higher order uncertainties. However, if other systematic uncertainties were added to the LECs in \cite{Colangelo:2001df} our agreement with them would be even better. Finally, let us emphasize the differences between our approach and that in \cite{Colangelo:2001df}. We are obtaining our $b_i$ from a fit to threshold parameters up to third order, using sum rules calculated with a fit to data constrained by dispersion relations. Note that the dispersive constraints are not imposed exactly, but only within experimental uncertainties \cite{GarciaMartin:2011cn}. Also, we are including the very precise NA48/2 data and we do not impose any constraint from chiral symmetry. In contrast, in \cite{Colangelo:2001df}, the dispersion relations are solved exactly, producing solutions parametrized in terms of the scattering phase at 800 MeV and the scattering lengths constrained by ChPT. Therefore, no fit below 800 MeV is performed  in \cite{Colangelo:2001df}, and the experimental input comes from energies above 800 MeV, or waves with angular momentum larger than 1. Thus, ours is a pure data analysis, including the most recent set. Our sum rules are largely dominated by these data below 800 MeV, and therefore we are showing that the data are very consistent with the two-loop ChPT representation of threshold parameters, with the exception of $c_P$. Also, the ChPT parameters we obtain by excluding $c_P$ are very consistent with those determined in \cite{Colangelo:2001df} using sum rules with input from Roy equations and two-loop ChPT constraints, but without fitting the data below 800 MeV.

Actually, the mismatch between the sum rule result for $c_P$
 and its one- and two-loop  calculations, 
had already been observed in \cite{AB2} using a sum rule which,
as we will show in Appendix~\ref{ap:sum rules}, corresponds to approximating
the absorptive parts inside the integral by the $S$ and $P$ waves only. In addition
they neglected all absorptive contributions above the two-kaon threshold. 
 Like us, the authors attributed this mismatch between ChPT and the sum rule result,
 to the presence of the $\rho(770)$ in this partial wave, which seems to require higher order corrections in ChPT.
Similarly to us, they found a reasonable agreement with the other threshold coefficients, 
but not for $c_P$,
whose standard two-loop ChPT calculation was almost a factor of 3 smaller than their sum rule result. Furthermore, a similar one-loop calculation even had the sign wrong.
Unfortunately they did not provide uncertainties in their calculation, but just
several values for different input,
so it was not straightforward to estimate the significance of such a mismatch. Their two-loop calculation is exactly the same as our rounded central value.
In contrast, for the sum rule we obtain a 40\% higher result, but one should take into account that they only used $S$ and $P$ waves as input, and only up to two-kaon threshold, plus some crude estimate of the $f_2$ contribution, whereas we use input up to $F$ waves and tens of GeV, as well as the latest NA48/2 data at lowest energy, which were not available then. As can be noticed from
Table \ref{tab:our-estimate}, we confirm the existence of such a mismatch, although is slightly lower than
the one already observed, since our final result for $c_P$ is slightly smaller than that obtained in \cite{AB2}. Moreover, 
taking into account our estimated uncertainties, the mismatch between the sum rule $c_P$ parameter and the ChPT result is of the order of 3 or 4 deviations, depending on whether one prefers to add the uncertainties in quadrature or linearly, given that they are largely of a systematic character.

To conclude, the values obtained for the threshold parameters using this averaged set of LECs
are shown in the second column of Table~\ref{tab:our-estimate}, where we can see
that, with the exception of $c_P$, they are rather compatible with the experimental
determination. Being also in quite good agreement with existing determinations, it is not surprising that our ``Estimates $O(p^6)$'' set satisfies well the existing axiomatic constraints that contain also $O(p^6)$ LECs \cite{axiomatic-bs}.

\section{Summary and Discussion}
\label{sec:conclusions}

In this work, we have determined the low energy constants of $SU(2)$ Chiral Perturbation Theory (ChPT) at one and two loops from a fit to the threshold parameters that were obtained from sum rules using a recent and precise dispersive analysis of data~\cite{GarciaMartin:2011cn}, together with six additional observables that we have studied here. 

Threshold parameters are defined as the coefficients of the effective range expansion of $\pi\pi$ scattering partial waves, which in this work we have studied up to angular momentum $\ell=3$, i.e., $S$, $P$, $D$ and $F$ waves, and all possible isospin states $I=0,1,2$. The coefficients of the two first orders, namely
the scattering lengths $a_{\ell I}$ and slope parameters $b_{\ell I}$, were already obtained from a dispersive analysis of data in \cite{GarciaMartin:2011cn}. In addition, we have provided here three sum rules to estimate the third order coefficients $c_{\ell I}$, thus adding six new observables to form a total set of 18.
For completeness, we have provided an appendix with a compilation
of all the sum rules used in this work, whether they have been derived here or not. 
Moreover, we have briefly reviewed how the different terms and low energy constants of ChPT contribute to each one of these threshold parameters and we have explained how the sum rules for the $c$ coefficients are related to previous results in the literature when approximating the absorptive parts in the integrals just by the $S$ and $P$ wave contributions.
                
We have then proceeded to fit these observables, first within one-loop ChPT, $O(p^4)$, and then to the full two-loop $O(p^6)$ calculation \cite{Bijnens:1995yn}. We have checked that the one-loop formalism is  clearly insufficient to accommodate the present level of precision. There is a clear improvement, in terms of $\chi^2/d.o.f.$ when using the two-loop expansion, although it is still not sufficient to get a good quality fit. This suggests that even higher order ChPT contributions may still be required to describe all these observables simultaneously.

However, we have been able to identify that the largest incompatibility is due to the $c_P$ parameter, confirming earlier findings \cite{AB2}. This may not come as a big surprise, since the largest contribution to the value of the sum rule that determines $c_P$ is given by the $\rho$ resonance and its sharp rise before 770 MeV,
which cannot be reproduced by the perturbative ChPT series.
Actually, we have estimated, by changing $f_\pi$ from its physical value to its value in the chiral limit in the last term of the $O(p^6)$ expansion, that this observable is a natural candidate to receive very large corrections from higher ChPT orders.

Hence, if $c_P$ is omitted, the quality of the two-loop fit improves, although there is still some tension
in the parameters to describe the remaining threshold parameters. Nevertheless, by conveniently
using $f_0$ in the last term of the ChPT expansion, the two-loop expansion can provide an acceptable description
of the rest of the threshold observables. The ChPT parameters  thus obtained, for which we provide statistical errors as well as an estimate of systematic uncertainties, are fairly compatible with previous determinations. We hope that the precise low energy constants determined in this work, together with their estimated uncertainties, can be of use for future studies of ChPT.

\section*{Acknowledgments} J.R.P. thanks the late F. J. Yndur\'ain for many discussions about the possibility to carry out this project after completing \cite{GarciaMartin:2011cn}. J. N. acknowledges funding by the German Academic Exchange Service (DAAD) and the Fundaci\'on Ram\'on Areces. G. R.	acknowledges	partial support by the EU Integrated Infrastructure Initiative HadronPhysics3 Project under Grant No. 283286 and by DFG (CRC 16, ÔÔSubnuclear Structure of MatterÕÕ and CRC 110, ÔÔSymmetries and the Emergence of Structure in QCDÕÕ)
We also thank B. Kubis and U. -G. Mei{\ss}ner for their reading of the manuscript and their valuable suggestions and corrections, Z. H. Guo and H. Q. Zheng for clarifying the use of their axiomatic bounds and B. Ananthanarayan for kindly helping us with the existing sum rules for $c$ parameters.
This work is partly
supported by DGICYT contracts FPA2011-27853-C02-02, FPA2010-17806, and the EU Integrated
Infrastructure Initiative Hadron Physics Project under contract
RII3-CT-2004-506078.

\appendix
\section{Fits to $\bar l_i$ and $\tilde r_i$}
\label{subsec:ls&rs}

\begin{table*}[b]
 \centering
 \footnotesize
    \renewcommand{\arraystretch}{1.3}
 \begin{tabular*}{\textwidth}{p{2.85cm}ccccccccccc}
   \hline \hline
  Fit to: & $\bar l_1$ & $\bar l_2$& $\bar l_3$& $\bar l_4$& $\tilde r_1$&
$\tilde r_2$& $\tilde r_3$& $\tilde r_4$& $\tilde r_5$& $\tilde r_6$&
 $\chi^2/d.o.f.$ \\
   \hline
   All &  -0.88$\pm$1.43 &  5.1$\pm$ 0.8 &  -49$\pm$10 &  4.5$\pm$1.3
&  -984$\pm$335 &  -101$\pm$302 &    -5.7$\pm$26 &   -13$\pm$15 &
1.6$\pm$0.9 &  0.45$\pm$0.33 &  $\frac{42}{18-10+1}$=5  \\
All but $c_P$ &  -2.2$\pm$1.5 &  5.6$\pm$0.8 &  -20$\pm$11 & 10$\pm$2 &
276$\pm$845 & -1361$\pm$549 &    34$\pm$36 &   -38$\pm$19 &
0.67$\pm$0.94 &  0.66$\pm$0.34 &  $\frac{6.7}{17-10+1}$=0.8 \\
All but $c_P$ with $f_0$ &  -0.5$\pm$1.0 &  4.2$\pm$0.6 &   -6$\pm$8 &
6.6$\pm$1.1 &    46$\pm$450 &  -356$\pm$238 &     4$\pm$15 &
-9$\pm$9 &  1.5$\pm$0.6 &  0.5$\pm$0.2 &   $\frac{4.7}{17-10+1}$=0.6 \\
   \hline
   Constrained fit to: &  \\
   \hline
All but $c_P$
 &  -0.11$\pm$0.16 &  4.2$\pm$0.1 &  3.3 &  5.8$\pm$0.4 &
-1.5 &     3.2 &    -4.2 &    -2.5 &
3.1$\pm$0.5 &  0.85$\pm$0.15 &  $\frac{68}{17-5+1}$=5.2\\
All but $c_P$ with $f_0$ &  0.5$\pm$0.2 &  3.9$\pm$0.1 &
3.3 &  5.1$\pm$0.3 &    -1.5 &     3.2 &
-4.2 &    -2.5 &  1.4$\pm$0.4 &  0.47$\pm$0.12 &
$\frac{15.7}{17-5+1}$=1.2 \\
   \hline
   \hline
{\bf    Our Estimate $O(p^6)$ } &{\bf  0.4$\pm$0.5 }&{\bf  3.9$\pm$0.3 }&{\bf  3.3 }&{\bf 5.2$\pm$0.7 }& {\bf    -1.5  }& {\bf     3.2  }& {\bf -4.2  }& {\bf    -2.5  }& {\bf 1.7$\pm$1.5  }& {\bf 0.5$\pm$0.3  }& ---\\
  Our Estimate $O(p^4)$ & -1.5$\pm$0.5 & 5.3$\pm$0.7 & -3$\pm$7 &
6.0$\pm$1.2 & --- & --- & --- & --- & --- & --- & ---   \\
   \hline
   \hline
Ref.~\cite{Bijnens:2011tb} ``All NLO''& 1.1 & 4.6 & 4.9 & 4.8 & --- & --- & --- & --- & --- & --- & --- \\
Ref.~\cite{Bijnens:2011tb} ``All NNLO''& -0.1 & 5.3 & 4.2 & 4.8 & --- & --- & --- & --- & --- & --- & --- \\
       Refs.\cite{Bijnens:1995yn,Colangelo:2001df,Colangelo:2010et} & -0.4$\pm$0.6 & 4.3$\pm$0.1 & 3.3$\pm$0.7 & 4.4$\pm$0.2 &
   -1.5 & 3.2 & -4.2 & -2.5 & 3.8$\pm$1.1 & 1.0$\pm$0.1 & --- 
 \end{tabular*}
       \caption{$O(p^6)$ fits to different sets of threshold parameters using the low energy constants $\bar l_i$ and $\tilde r_i$. In the first row we fit all the
    threshold parameters obtained in this work. Note that the $\chi^2/d.o.f.$ is quite large. However, as it happened in the equivalent fit in Sec.~\ref{sec:op6fits}, the largest contribution to $\chi^2$ comes from $c_P$. Thus, in the following fits we include all the observables but $c_P$. In the second and third rows we show the LECs obtained when $c_P$ is excluded, using $f_\pi$ and $f_0$ in the last term of the ChPT expansion, respectively. The quality of the fits notably improves. Nevertheless, the large size of the errors in the case of $\bar l_3$, $\tilde r_1$, $\tilde r_2$, $\tilde r_3$ and $\tilde r_4$ indicates that our fits are not very sensitive to these LECs. For that reason, in the next section of the table (``Constrained fit to'') we repeat the latter two fits, fixing the value of $\bar l_3$ to an average of  lattice determinations~\cite{Colangelo:2010et} and that of the LECs $\tilde r_1$ to $\tilde r_4$ to the resonance saturation estimates~\cite{Bijnens:1995yn}. We provide an estimate of the LECs and their uncertainties as a weighted average of these two last fits. Let us note that the $O(p^4)$ LECs $\bar l_1$ to $\bar l_4$ do not lie too far from our $O(p^4)$ estimates, shown immediately below. Moreover, within our uncertainties, the resulting values are very consistent with previous determinations, particularly with those listed in the last row ($\bar l_3$ from~\cite{Colangelo:2010et}, $\tilde r_1$ to $\tilde r_4$ from~\cite{Bijnens:1995yn} and the rest of the LECs from~\cite{Colangelo:2001df}), remembering that the latter only include the ``noise'' in their evaluations and not systematic uncertainties.}
   \label{tab:ls&rs}
 \end{table*}
 
 As commented in Sec.~\ref{sec:op6fits}, the two-loop $\pi\pi$
scattering amplitudes can be recast in terms of six independent terms
multiplied by their corresponding low energy constants $\bar b_i$.
In turn, these $\bar b_i$ can be rewritten in terms of the 
four
$O(p^4)$ LECs that appear in the Lagrangian and six combinations $r_i$ of the $O(p^6)$ LECs. The difference between writing the amplitude in one way or the other is $O(p^8)$.
However, despite increasing the number of parameters to ten, 
the $O(p^6)$ amplitude still provides just six independent structures.
As a consequence, the fits in terms of  $\bar l_i$ and $ r_i$ are
 much more unstable, and can even lead to spurious solutions. 
 For this reason we have explained the fits in terms of $\bar b_i$ in the main text,
 and we have relegated the $\bar l_i$, $r_i$ fits to this appendix.
  
Let us then revisit the fits of Table~\ref{tab:fullbs} where we fit all 
the observables, or all but $c_P$, but recasting the 
amplitudes in terms of $\bar l_i$ and $\tilde r_i$ \footnote{We give the fit
results in terms of $\tilde r_i$, defined as $\tilde r_i=(4\pi)^4r_i$, which 
have a size of order one.}.
The resulting values are given in Table~\ref{tab:ls&rs}.
We observe the same pattern as before: the fit to all 
parameters still has a rather high $\chi^2/d.o.f.=5$,
because, although the total $\chi^2$ has decreased from 67 to 42, 
the number of degrees of freedom has increased by 4.
Let us remark that $\tilde r_1$ and $\tilde r_2$ have central 
values many orders of magnitude bigger than expected,
but their uncertainties are comparably large. 
This means that we do not have any real sensitivity to these parameters.

Once again, the largest contribution to the $\chi^2$ is due to $c_P$, and thus we
remove it from the fits, as we did in the main text. When so doing, the $\chi^2/d.o.f.$ becomes less than 1, yielding a statistically acceptable fit, but of course, the uncertainties are still huge for  $\tilde r_1$ and $\tilde r_2$ 
and very large for $\bar l_3$, $\tilde r_3$ and $\tilde r_4$. The central value of $\bar l_3$ is also
far from our $O(p^4)$ values, also given in Table~\ref{tab:ls&rs}, or the lattice value in \cite{Colangelo:2010et}. Similarly, those  of $ \tilde r_1$ to $\tilde r_4$ are far from the resonance saturation estimates in \cite{Bijnens:1995yn}. However, all of them are still relatively compatible due to the resulting large uncertainties.

As we did in the main text, we repeat this last fit to all parameters but $c_P$, replacing $f_\pi$ by $f_0$ in the $O(p^6)$ terms, which is a change of higher order in the ChPT expansion. We observe that we obtain a similarly good description of the observables, but with LECs closer to the reference values from~\cite{Bijnens:1995yn,Colangelo:2001df,Colangelo:2010et}.

In fact, since we observe that our fits are not very sensitive to the value of some LECs, we can ask what quality we can achieve if we fix these parameters to the reference values. We thus repeat the fit to all the observables but $c_P$, fixing $\bar l_3$ and $\tilde r_1$ to $\tilde r_4$ to the reference values given in the last row of Table~\ref{tab:ls&rs}, both using $f_\pi$ and $f_0$ in the $O(p^6)$ terms. The results of these ``constrained'' fits are shown in the fourth and fifth rows of the table. As expected, the $\chi^2/d.o.f.$ does not increase much with respect to the unconstrained fit, confirming that our observables depend little on these LECs. Thus, as we did in the main text with the $O(p^4)$ and the $O(p^6)$ $b_i$ parameters, we provide in Table~\ref{tab:ls&rs} here an ``Estimated $O(p^6)$'' set as the weighted average of the values found in the constrained fits leaving the $\bar l_3, \tilde r_1,\tilde r_2,\tilde r_3$ and $\tilde r_4$ fixed. These estimates come fairly compatible within uncertainties with the values existing in the literature \cite{Bijnens:1995yn,Colangelo:2001df,Colangelo:2010et,Bijnens:2011tb}, particularly with those in \cite{Bijnens:1995yn,Colangelo:2001df,Colangelo:2010et}. Thus, it comes as no surprise that they also satisfy quite comfortably the $O(p^6)$ constraints \cite{axiomatic-bs}, as it already happened with our $b_i$ in the main text. Concerning the axiomatic bounds in \cite{axiomatic-ls-and-rs}, we should recall that they are derived in the large $N_c$ limit, and therefore they cannot be directly compared with sets obtained in the physical regime, i.e. $N_c=3$, like ours or those in \cite{Colangelo:2001df}. Actually, if applied blindly to either our set or that in \cite{Colangelo:2001df}, the bounds would be violated. The leading $1/N_c$ part has to be extracted, but this requires additional theoretical input and assumptions beyond our present scope.

\section{Threshold parameters in ChPT}

We show here the two-loop ChPT expressions for the
$F$-wave threshold parameters as well as the third order threshold
parameters, $c_{\ell I}$, for all waves. The scattering lengths, $a_{\ell I}$, 
and slope parameters, $b_{\ell I}$,
for the $S$, $P$ and $D$ waves can be found, for instance, in~\cite{Bijnens:1995yn}.

\begin{widetext}
 \begin{eqnarray}
   \label{threshols_pars_chPT_F}
    \begin{aligned}
      &a_{F}=\frac{11}{94080\pi^3f_\pi^4M_\pi^3}
      \left[
        1+\frac{M_\pi^2}{132\pi^2f_\pi^2}
        \left(
          9 \bar b_1+\frac{51}{2} \bar b_2-151 \bar b_3-653 \bar b_4+126 \bar b_5
          +126 \bar b_6+\frac{111 \pi^2}{20}+\frac{4111}{9}
        \right)
      \right],
      \\[9pt]
      &b_{F}=-\frac{47}{529200\pi^3f_\pi^4M_\pi^5}
      \left[
        1+\frac{M_\pi^2}{752\pi^2f_\pi^2}
        \left(
          75 \bar b_1+169 \bar b_2+2874 \bar b_3+6270 \bar b_4
          +\frac{205 \pi ^2}{6}-\frac{549221}{360}
        \right)
      \right],
      \\[9pt]
      &c_{F}=\frac{463}{3274425\pi^3f_\pi^4M_\pi^7}
      \left[
        1+\frac{5M_\pi^2}{29632\pi^2f_\pi^2}
        \left(
          675 \bar b_1+\frac{7079}{5} \bar b_2+14341 \bar b_3+
          \frac{134517}{5} \bar b_4+200 \pi ^2+\frac{2301641}{900}
      \right)
    \right].
      \\[9pt]
  \end{aligned}
\end{eqnarray}
\end{widetext}

\begin{widetext}
 \begin{eqnarray}
   \label{threshols_pars_chPT}
    \begin{aligned}
      &\begin{split}c_{S0}=
        \frac{1}{3456 \pi^3 f_\pi^4M_\pi}
        \bigg[
        &792\bar b_3+1224\bar b_4 - 253\\
        &+\frac{M_\pi^2}{\pi^2f_\pi^2}
        \left(
          -\frac{219}{8}\bar b_1-59 \bar b_2+\frac{2893}{5} 
          \bar b_3+\frac{4781}{5} \bar b_4+486 \bar b_5-90
          \bar b_6-\frac{91589 \pi ^2}{384}+\frac{685061}{480}
        \right)
        \bigg],
      \end{split}
      \\[9pt]
      &c_{S2}=\frac{1}{8640\pi^3 f_\pi^4M_\pi}
      \bigg[
      360 \bar b_3+2520 \bar b_4-19
      +\frac{M_\pi^2}{\pi^2f_\pi^2}
      \left(
        \frac{267}{8} \bar b_1+31 \bar b_2-106 \bar b_3-956 \bar b_4
        -360 \bar b_6+\frac{1049 \pi ^2}{12}-\frac{75997}{96}
      \right)
      \bigg],
      \\[9pt]
      &c_{P}=\frac{-23}{6720\pi^3 f_\pi^4M_\pi^3}
        \bigg[
        1-\frac{M_\pi^2}{3726\pi^2f_\pi^2}
        \bigg(
        \frac{729}{2} \bar b_1-\frac{405}{4} \bar b_2-\frac{23823}{2} \bar b_3
        -\frac{36261}{2} \bar b_4+5103(\bar b_5 + \bar b_6)
        +\frac{1551 \pi ^2}{40}+21037
        \bigg)
        \bigg],
      \\[9pt]
      &c_{D0}=\frac{499}{264600\pi^3f_\pi^4M_\pi^5}
      \left[
        1+\frac{M_\pi^2}{17964\pi^2f_\pi^2}
        \left(
          135 \bar b_1+\frac{4761}{2} \bar b_2+19701 \bar b_3-29079 \bar b_4
          +\frac{2667 \pi ^2}{4}-\frac{18331}{10}
        \right)
      \right],
      \\[9pt]
      &c_{D2}=\frac{127}{105840\pi^3f_\pi^4M_\pi^5}
      \left[
        1+\frac{M_\pi^2}{2540\pi^2f_\pi^2}
        \left(
          102 \bar b_1+\frac{839}{2} \bar b_2+5053 \bar b_3+5421 \bar b_4
          -\frac{73 \pi ^2}{12}-\frac{342941}{360}
        \right)
      \right],
      \\[9pt]
  \end{aligned}
\end{eqnarray}
\end{widetext}

\section{Sum rules}
\label{ap:sum rules}

As described in the main text, we have used sum rules, derived from forward dispersion relations (FDRs) or from the Froissart-Gribov representation. Some of them have been derived in this work and explained in the main text, and some others can be found in the literature, and we just provide them here for completeness and convenience. This is not supposed to be a review, we only list here those sum rules that we have actually used in our calculations.

\subsection{Sum rules from Forward Dispersion Relations}

We start with those sum rules obtained from FDRs. First of all we recall the classic Olsson sum rule \cite{Olsson}, obtained at threshold from an FDR for the antisymmetric $F^{I_t=1}$ amplitude:
\begin{equation}
  2a_0^{0}-5a_0^{2}= 3M_\pi\int_{4M_\pi^2}^\infty \dd\! s\,
\dfrac{\imag F^{I_t=1}(s,0)}{s(s-s_{th})},
\label{eq:OlssonSR}
\end{equation}
where $s_{th}=4M_\pi^2$.

Next we list those obtained for the slope and shape parameters, which read
\begin{widetext}
\begin{align}
  \label{eq:sumrules}
  b_{S0}&=\frac1{2M_\pi}\lim_{s\to\sth^+}
  \left(
    {\rm P.P.}\int_{\sth}^{\infty}ds'
    \left[
      \frac{\im F^0(s',0)}{(s'-\sth)(s'-s)}-
      \frac{\im F^0(s',0)-3\im F^1(s',0)+5\im F^2(s',0)}{3s'(s'+s-\sth)}
    \right]
  \right),\\
  c_{S0}&=-10a_2^0+\frac{4}{M_\pi}\,\frac{d}{ds}
  \left(
    {\rm P.P.}\int_{\sth}^\infty ds'
    \left[
      \frac{\im F^0(s',0)}{(s'-\sth)(s'-s)}-
      \frac{\im F^0(s',0)-3\im F^1(s',0)+5\im F^2(s',0)}{3s'(s'+s-\sth)}
    \right]
  \right)_{s=\sth^+},\\
  b_{S2}&=\frac1{2M_\pi}\lim_{s\to\sth^+}
  \left(
    {\rm P.P.}\int_{\sth}^{\infty}ds'
    \left[
      \frac{\im F^1(s',0)}{(s'-\sth)(s'-s)}-
      \frac{2\im F^0(s',0)+3\im F^1(s',0)+\im F^2(s',0)}{6s'(s'+s-\sth)}
    \right]
  \right),\\
  c_{S2}&=-10a_2^2+\frac{4}{M_\pi}\,\frac{d}{ds}
  \left(
    {\rm P.P.}\int_{\sth}^\infty ds'
    \left[
    \frac{\im F^1(s',0)}{(s'-\sth)(s'-s)}-
    \frac{2\im F^0(s',0)+3\im F^1(s',0)+\im F^2(s',0)}{6s'(s'+s-\sth)}
    \right]
  \right)_{s=\sth^+},\\
  b_P&=\frac1{3M_\pi}\,\frac{d}{ds}
  \left(
    {\rm P.P.}\int_{\sth}^\infty ds'
    \left[
      \frac{\im F^1(s',0)}{(s'-\sth)(s'-s)}+
      \frac{2\im F^0(s',0)-3\im F^1(s',0)-5\im F^2(s',0)}{6s'(s'+s-\sth)}
    \right]
  \right)_{s=\sth^+},\\
  c_P&=-\tfrac{14}{3}a^1_3+\frac{8}{3M_\pi}\,\frac{d^2}{ds^2}
  \left(
    {\rm P.P.}\int_{\sth}^\infty ds'
    \left[
      \frac{\im F^1(s',0)}{(s'-\sth)(s'-s)}+
      \frac{2\im F^0(s',0)-3\im F^1(s',0)-5\im F^2(s',0)}{6s'(s'+s-\sth)}
    \right]
  \right)_{s=\sth^+}.
\end{align}
\end{widetext}
Those for the $b$ parameters were already provided by one of us in \cite{PY05,KPY08}.
Note that they correspond to the threshold limit, taken from above, of the appropriate FDR for each channel, or to its first or second derivative. As explained in the main text, the
appropriate FDR can be obtained as a combination of FDRs for the  $s\leftrightarrow u$ symmetric $F^{0+}$,  $F^{00}$ and antisymmetric $F^{I_t=1}$ amplitudes. Note that the latter only needs one subtraction. Here we have chosen to write them in terms of $F^I$ amplitudes, which are more convenient for calculations in terms of partial waves. The other version is more convenient for Regge theory expressions and for the derivation itself.

We can obtain a more usable expression for the sum rules by
removing the principal parts (${\rm P.P.}$) of the integrals. First note that
the ${\rm P.P.}$ only affects the first term of the integrals, the one with the
pole at $s'=s$. A way to remove the ${\rm P.P.}$ is to subtract zero, 
which we have done by writing
\begin{align}
  \label{eq:PV-zero}
  \underbrace{\frac{\im F^I(s,0)}{\sqrt{s-\sth}}}_{\equiv g_I(s)}\,
  P.P.\int_{\sth}^{\infty}\frac{ds'}{\sqrt{s'-\sth}(s'-s)}=0,
\end{align}
so that we do not change the value of the integral but this new piece cancels
the pole at $s'=s$, so the ${\rm P.P.}$ divergence disappears and we can easily evaluate
the limits and derivatives. Then we obtain
\begin{widetext}
\begin{align}
  \label{eq:sumrulesPP} 
  b_{S0}&=\frac1{2M_\pi}\int_{\sth}^\infty ds
  \left[
    \frac{\im F^0(s,0)}{(s-\sth)^2}-\frac{g_0(\sth)}{(s-\sth)^{3/2}}
    -\frac{\im F^0(s,0)-3\im F^1(s,0)+5\im F^2(s,0)}{3s^2}
  \right],
\\
   c_{S0}&=-10a_{D0}+\frac{4}{M_\pi}\int_{\sth}^\infty ds
   \left[
     \frac{\im F^0(s,0)}{(s-\sth)^3}-\frac{g_0(\sth)+g_0'(\sth)(s-\sth)}{(s-\sth)^{5/2}}
     +\frac{\im F^0(s,0)-3\im F^1(s,0)+5\im F^2(s,0)}{3s^3}
   \right],
\\
   b_{S2}&=\frac1{2M_\pi}\int_{\sth}^\infty ds
   \left[
     \frac{\im F^2(s,0)}{(s-\sth)^2}-\frac{g_2(\sth)}{(s-\sth)^{3/2}}
     -\frac{2\im F^0(s,0)+3\im F^1(s,0)+\im F^2(s,0)}{6s^2}
   \right],
\\
   c_{S2}&=-10a_{D2}+\frac{4}{M_\pi}\int_{\sth}^\infty ds
   \left[
     \frac{\im F^2(s,0)}{(s-\sth)^3}-\frac{g_2(\sth)+g_2'(\sth)(s-\sth)}{(s-\sth)^{5/2}}
     +\frac{2\im F^0(s,0)+3\im F^1(s,0)+\im F^2(s,0)}{6s^3}
   \right],
\\
   b_P&=\frac1{3M_\pi}\int_{\sth}^\infty ds
   \left[
     \frac{\im F^1(s,0)}{(s-\sth)^3}-\frac{2\im F^0(s,0)-3\im F^1(s,0)-5\im F^2(s,0)}{6s^3}
   \right],
\\
   c_P&=-\tfrac{14}{3}a_F+\frac{8}{3M_\pi}\int_{\sth}^\infty ds
   \left[
     \frac{2\im F^1(s,0)}{(s-\sth)^4}-\frac{g_1''(\sth)}{(s-\sth)^{3/2}}
     +\frac{2\im F^0(s,0)-3\im F^1(s,0)-5\im F^2(s,0)}{3s^4}
   \right],
\end{align}
\end{widetext}
where $g_I(s)=\im F^I(s,0)/\sqrt{s-\sth}$, the primes denote derivatives with
respect to $s$, and we have used that $g_1(\sth)=g_1'(\sth)=0$. 
In the main text, Eqs.\eqref{eq:newSRcp}-\eqref{eq:newSRcS0}, we have recast the formulas in terms of the symmetric $F^{0+}$ and $F^{00}$ amplitudes and  we have further developed the expressions of the sum rules for $c_i$ by 
writing explicitly the content of $g_I(s)$:
\begin{align}
  \label{g_I}
  \nonumber
  &g_I(s)=\frac{M_\pi}{\pi}\bigg[4a_{SI}^2
+\left(
    a_{SI}^4+2a_{SI}b_{SI}-\frac{a_{SI}^2}{2 M_\pi^2}
  \right)(s-\sth)
  \\
  &\qquad \qquad \qquad +O(s-\sth)^2\bigg],\quad I=0,2,
  \\
  &g_1(s)=\frac{3 M_\pi}{4\pi}a_P^2(s-\sth)^2+O(s-\sth)^3.  
\end{align}

\subsection{Sum rules from the Froissart-Gribov representation}
\label{appendix:FG}

Next, for $\ell\geq1$, we list the sum rules \cite{Palou:1974ma} obtained from the Froissart-Gribov representation \cite{FG,Yndurain:1972ix} of the $t$ channel partial wave expansion (see \cite{Yndurain:2002ud} for a pedagogical review). In particular we used the antisymmetric $F^{I_t=1}$, whose dispersion relation does not require subtractions, as well as the two symmetric $F^{0+}$ and $F^{00}$ waves, which do need one. Let us note that, irrespective of the number of subtractions in the $t$ channel dispersion relation for these three amplitudes, the resulting form of the sum rule reads the same, as long as  $\ell\geq1$. These sum rules read:
\begin{widetext}
\begin{eqnarray}\label{eq:apppendixFG}
a_\ell&=&\,\dfrac{\sqrt{\pi}\,\gammav(\ell+1)}{4M_{\pi}\gammav(\ell+3/2)}
\!\int_{4M_{\pi}^2}^\infty ds\,\dfrac{\imag F(s,4M_{\pi}^2)}{s^{l+1}},
\\ \nonumber
b_\ell&=&\,\dfrac{\sqrt{\pi}\,\gammav(\ell+1)}{2M_{\pi}\gammav(\ell+3/2)}
\!\int_{4M_{\pi}^2}^\infty ds\,
\Big\{\dfrac{4\imag F'_{\cos\theta}(s,4M_{\pi}^2)}{(s-4M_{\pi}^2)s^{\ell+1}}-
\dfrac{(\ell+1)\imag F(s,4M_{\pi}^2)}{s^{\ell+2}}\Big\},\hspace{-.5cm}
\\ \nonumber
 c_{\ell } &=& \frac{\sqrt{\pi}\,\Gamma(\ell+1)}{M_\pi\,\Gamma(\ell+3/2)}
\!\int_{4M_\pi^2}^{\infty} \!\!\!\!ds
\left\{\frac{16 \,{\im F}''_{\cos\theta}(s,4M_\pi^2)}{(s-4M_\pi^2)^2 s^{\ell+1}}
\!-\!8(\ell+1)\frac{\im {F}'_{\cos\theta}(s,4M_\pi^2)}{(s-4M_\pi^2)s^{\ell+2}}
+\frac{\im F(s,4M_\pi^2)}{s^{\ell+3}} \frac{(\ell+2)^2(\ell+1)}{\ell+3/2}\right\},
\end{eqnarray}
\end{widetext}
where now $F$ stands for $F^{0+},F^{00}$ or $F^{I_t=1}$, but we have suppressed that label to simplify the notation. Hence, the $I=1$ threshold parameters come directly out of the  $F^{I_t=1}$
sum rules, whereas for the $I=0,2$ ones, coming from $D$ waves, 
one has to recall that: $a^{00}=2(a_{D0}/3+2a_{D2}/3)$ and $a^{0+}=2(a_{D0}/3-a_{D2}/3)$. Note also that we have defined  $\imag F'_{\cos\theta}\equiv (\partial/\partial\cos\theta_s)\imag F$, where $\cos\theta_s$
is the angle between the initial and final pions. 
Please note that for amplitudes with fixed isospin in the $t$ channel, an extra factor of 2 
(due to identity of particles) has to be added to the left-hand side of the equation above (see, for instance the explicit formulas in \cite{PY05}).

\subsection{Comparison with the sum Rules in Ref.~\cite{AB2}}

We are not the first ones to derive sum rules for third order threshold parameters. 
Actually, in Ref.~\cite{AB2} sum rules for shape parameters were already obtained
from Roy equations in the $S$- and $P$-wave approximation for the absorptive parts
inside the integrals. 
For completeness, we reproduce those sum rules here, correcting some typos 
\footnote{We thank B. Ananthanarayan for his kind help confirming these expressions.}.
Please note a factor of 2 difference between the 
coefficients of \cite{AB2}, which we denote by $c^{AB}$, and our definition here, namely $c=2 c^{AB}$. 
\begin{eqnarray}
  \label{c00-anan}
  \nonumber
  c_{S0}&=&\dfrac{128}{\pi}\int_0^{\infty}d\nu
  \bigg[
  \frac{5\,\im t_0^2(\nu)}{288(1+\nu)^3}-\frac{(1+2\nu)\im t_1^1(\nu)}{32\nu(1+\nu)^3}
  \\\nonumber
  &&+\left(
    \frac{1}{64\nu^3}+\frac1{288(1+\nu)^3}
  \right)\im t_0^0(\nu)
  \\\nonumber
  &&-\frac{(1+2\nu)\sqrt{\nu(\nu+1)}}{256\pi\nu^3(1+\nu)^3}
  \Big\{
  \sigma^0(0)(1+\nu+\nu^2)
  \\
  &&+(\nu+\nu^2)\frac{d\sigma^0(\nu)}{d\nu}\vert_{\nu=0}
  \Big\}
  \bigg],
\end{eqnarray}

\begin{eqnarray}
  \label{c20-anan}
  \nonumber
  c_{S2}&=&\frac{128}{\pi}\int_0^{\infty}d\nu
  \bigg[
  \frac{\im t_0^0(\nu)}{288(1+\nu)^3}+\frac{(1+2\nu)\im t_1^1(\nu)}{64\nu(1+\nu)^3}
  \\\nonumber
  &&+\left(
    \frac{1}{64\nu^3}+\frac1{576(1+\nu)^3}
  \right)\im t_0^2(\nu)
  \\\nonumber
  &&-\frac{(1+2\nu)\sqrt{\nu(\nu+1)}}{256\pi\nu^3(1+\nu)^3}
  \Big\{
  \sigma^2(0)(1+\nu+\nu^2)
  \\
  &&+(\nu+\nu^2)\frac{d\sigma^2(\nu)}{d\nu}\vert_{\nu=0}
  \Big\}
  \bigg],
\end{eqnarray}

\begin{eqnarray}
  \nonumber
  \label{c11-anan}
  c_{P}&=&\frac{512}{\pi}\int_0^{\infty}d\nu
  \bigg[
  \frac{\im t_0^0(\nu)}{2560(1+\nu)^4}-\frac{\im t_0^2(\nu)}{1024(1+\nu)^4}
  \\\nonumber
  &&+\left(
    \frac{1}{256\nu^4}-\frac{2+11\nu}{5120\nu(1+\nu)^4}
  \right)\im t_1^1(\nu)
  \\
  &&-\frac{(1+2\nu)\sqrt{\nu(\nu+1)}}{6144\pi\nu^2(1+\nu)^2}
  )\frac{d^2\sigma^1(\nu)}{d\nu^2}\vert_{\nu=0}
  \bigg],
\end{eqnarray}
where $\nu=p^2/M_\pi^2$. Note that in order to get rid of the principal parts that would appear
otherwise in the integrals, some terms proportional to 
\begin{equation}
  \label{sigma}
\sigma^I(\nu)=\frac{4\pi}{\sqrt{\nu(\nu+1)}}\sum (2\ell+1)\im t_\ell^I(\nu),  
\end{equation}
or their derivatives, have been introduced in the integrals.

Now, the above sum rules for the $c$ coefficients were obtained in \cite{AB2}
from the Roy representation restricted to the $S$ and $P$ approximation for the absorptive parts in the integrals. Actually, if we also restrict our sum rules to that approximation and in addition we reexpress the $D$ and $F$ wave threshold parameters in terms of the Froissart-Gribov sum rules in Eq.\eqref{eq:apppendixFG} (beware the fact that the absorptive parts in the latter
 are evaluated at $t=4M_\pi^2$), we would recover the sum rules in Eqs.\eqref{c00-anan}, \eqref{c20-anan} and \eqref{c11-anan} above  {\it before the principal parts are removed}.
This was to be expected \cite{Mahoux:1974ej}, due to retaining only the $S$ and $P$ waves in the absorptive parts.
In \cite{AB2} they remove the principal part by
subtracting zero recast as
\begin{eqnarray}
0&=&\!\!\frac{s\,\im F^I(s)}{\sqrt{s(s-4)}}\,P.P.\!\!\int_{4}^\infty \!\!\!\!\!ds'
\frac{2s'-4}{\sqrt{s'(s'-4)}(s'+s-4)(s'-s)}\nonumber \\
&=&
\!\!P.P.\int_{4}^{\infty}  \!\!\!\!\!ds'
  \left[\frac{\im F^I(s)}{(s'-4)(s'-s)}+O(s'-s)^0\right],
\end{eqnarray}
where $\im F^I(s)/\sqrt{s(s-4)}\sim\sigma^I(s)$ and now $s$ is
given in $M_\pi$ units. Note that both
the equation above and Eq.~\eqref{eq:PV-zero} cancel the pole at
$s'=s$. Although their finite parts (the $O(s'-s)^0$ pieces) are
different, both integrals are zero, so they do not change the
value of the dispersion relation and one could use either one or the other
to remove the principal part. When taking into account these two different ways of 
removing the principal parts, we have checked that the sum rules in this
work reduce to those in \cite{AB2} in the $S$-$P$ approximation for the absorptive parts.
In this appendix, we have removed all principal parts using Eq.~\eqref{eq:PV-zero}
for consistency with the sum rules that we use to determine $a$ and $b$ parameters
from \cite{GarciaMartin:2011cn}. 

Of course, our sum rules are valid beyond the $S$-$P$ approximation.
In addition, our  evaluation of the $c$ parameters also differs from that in \cite{AB2}, because we are using a recent and precise dispersive determination of data, which includes the very relevant NA48/2 data near threshold. Actually, 
following the constrained fits in \cite{GarciaMartin:2011cn}, we use $S$, $P$, $D$ and $F$ waves up to 1420 MeV and
Regge fits to data beyond that energy. The uncertainties in these parametrizations also allow us to estimate the uncertainties in our sum rule calculation. As expected, the $D$, $F$ and high energy contributions are small, which can now be checked with real data parametrizations (the smallness of the $f_2(1275)$ contribution
was already checked in \cite{AB2}), but recall that we are obtaining very precise determinations of all the $c$ parameters and all these contributions must be kept under control. 

Numerically, the bulk of the result is given by the $S$ and $P$ contributions to the absorptive parts below 1 GeV, as was done in \cite{AB2}.
However, the rest of the contributions are sizable given our level of precision and cannot be neglected. In particular, 
if we take the absorptive parts from the dispersive
approach in  \cite{GarciaMartin:2011cn} but in
the $S$-$P$ approximation, also neglecting 
all their contributions beyond the two-kaon threshold, as done in \cite{AB2}, 
we would now obtain, either with our expressions or those in \cite{AB2}:
$c_{S2} \simeq(2.51\pm0.24)10^{-2}$ in pion mass units, consistent with the different sets
provided in \cite{AB2}, which lay in the range  $c_{S2}=(2.4\pm3.0)10^{-2}$. 
This is to be compared with our full result of $c_{S2}\simeq(2.79\pm0.24)10^{-2}$,
given in Table~\ref{abcexp} which includes the $D$ and $F$ waves and the higher energy contributions.
Concerning the $P$ wave, we would obtain $c_P=(1.51\pm0.14)10^{-3}$,
which is slightly lower than what was obtained in \cite{AB2}, which 
lay in the range $c_P=(1.6-1.9)10^{-3}$.
This is to be attributed to our using more recent parametrizations.
Also, this can be compared with our full result, given in Table~\ref{abcexp},
which yields $c_P=(1.35\pm0.15)10^{-3}$.
Finally, under the above approximations we would obtain $c_{S0}=(0.1\pm0.8)10^{-2}$, once again consistent with the several values given in \cite{AB2},
which lie on the range between $c_{S0}=-0.98\times10^{-2}$ and $1.64\times10^{-2}$. This can be directly compared to our full result for that sum rule of $c_{S0}=(0.7\pm0.8)10^{-2}$, given 
in Table~\ref{abcexp}, as well as to 
the final result once we average with the direct result, $c_{S0}=(0.45\pm0.67)10^{-2}$.
In summary, in all cases, the contribution from $D$ and $F$ waves and from energies higher than the two-kaon threshold amount to roughly 1 standard deviation of the total result.

\end{document}